\documentclass[twocolumn,aps,prl,groupedaddress,showpacs,slide,nofootinbib]{revtex4}

\usepackage{epsfig}
\usepackage{graphicx}
\usepackage{amssymb}
\usepackage{amsmath}
\usepackage{subfigure}
\usepackage{longtable}

\begin{document}

\title{Two Higgs Doublets Model  in Gauge-Higgs Unification framework}
\author{We-Fu Chang }
\email{wfchang@phys.nthu.edu.tw} 
\affiliation{Department of Physics, National Tsing Hua University, HsinChu 300, Taiwan}
\affiliation{Physics Division, National Center for Theoretical Sciences, HsinChu 300, Taiwan}
\author{Sin Kyu Kang}
\email{skkang@snut.ac.kr}
\affiliation{School of Liberal Arts, Seoul-Tech., Seoul 139-743, Korea}
\affiliation{Institute of Convergence Fundamental Studies, Seoul-Tech., Seoul 139-743, Korea}
 \author{ Jubin Park }
\email{honolo@phys.nthu.edu.tw}
\affiliation{Department of Physics, National Tsing Hua University, HsinChu 300, Taiwan}

\vspace{3cm}

\begin{abstract}
 We discuss the realization of two Higgs doublets model
in the framework of 6 dimensional Gauge-Higgs Unification model
with a simple Lie group $G_M$. Two Higgs $SU(2)_L$ doublets can
emerge at the low energy effective theory, and the quartic
coupling terms in the scalar potential,  essential for the
electroweak symmetry breaking, are now $G_M$ gauge invariant and
permissive. A realistic two Higgs doublets model can possibly be
obtained only when two of the root vectors associated with the
would-be Higgs doublets and the root vector for $SU(2)_L$ form an
isosceles triangle with vertex angle either of $\pi/3, \pi/2$, or
$2\pi/3$. Moreover, depending on $G_M$, the scalar potential of
resulting
 two Higgs doublets model can admit only a few limited forms. The mass spectrum of the
physical Higgs and the weak mixing angle are briefly discussed.

\end{abstract}


\maketitle

\section{Introduction}

 In the Gauge-Higgs Unification(GHU) models,
the vector fields of gauge group $G_M\supset  SU(2)_L\times
U(1)_Y$ propagate in  (4+d)-dimensional spacetime. The gauge
components in the $d$ compactified extra spatial dimensions behave
like scalar fields below  the compactification
scale~\cite{Manton:1979kb}. With properly chosen  gauge symmetry
and orbifolding boundary conditions(BC), an effective  scalar
$SU(2)_L$ doublet can  emerge at the low energy  and play the role
of Standard Model(SM) Higgs. Hence, we have no need of introducing
a fundamental scalar. Due to the higher dimensional gauge
symmetry, the $d$ extra scalar fields are massless.  The
spontaneous electroweak symmetry breaking(EWSB) in SM can be
triggered by the quantum corrections with the Wilson loop in the
non-simple connected space\cite{Hosotani:1988bm}. The notorious
gauge hierarchy problem associated with a fundamental Higgs boson
can be thus alleviated. For instance, a Higgs doublet could arise
from a 5 dimensional(5D) $SU(3)$ electroweak gauge theory on the
$S_1/Z_2$ orbifold\cite{1HD}. However, for $d=1$, the quartic
coupling term in the scalar potential must be generated by some
symmetry breaking quantum corrections for it vanishes at
tree-level as well. When $d\geq 2$, the quartic coupling terms in
the scalar potential, which arise from the square of field
strength, are gauge invariant by construction. Moreover, it is
possible to generate multi scalars at the low energy\cite{2HD}.

In this paper, we focus on the realization of two Higgs doublets
model (2HDM) in  6 dimensional(6D) GHU  models, bearing in mind
that (1)  2HDM predicts $\rho\equiv M_W^2/(M_Z^2
\cos^2\theta_W)=1$ after EWSB at tree-level, and (2) $d=2$ is the
minimal requirement to yield two Higgs ( not limited to $SU(2)_L$
doublets ).
 We shall exhaust all possible simple Lie groups
for $G_M$ and examine the resulting quartic coupling terms of
Higgs potential, denoted as $V_4$, which is  now completely
determined by group theory at tree-level. It is a delightful
surprise to us that $V_4$ can admit only a few forms for all
possible Lie groups. Our key finding is that, to successfully
generate a 2HDM at low energy, only the root vectors of $G_M$
associated with the would-be Higgs doublets and the root vector
for $SU(2)_L$ form an isosceles triangle with vertex angle either
of $\pi/3, \pi/2$, or $2\pi/3$. Moreover, $V_4$ solely depends on
the vertex angle. Our result is summarized in
Table-\ref{tab:table2}. On the other hand, the  quadratic terms of
Higgs potential, denoted as $V_2$, are assumed to be generated by
some symmetry breaking mechanism and they cannot be fixed by the
gauge symmetry. However, by using the physical Higgs mass
spectrum, one can parameterize  $V_2$ phenomenologically and
bypass the question of their origin. Finally, it is a well known
 difficulty  to construct a GHU model with the weak mixing
$\sin^2\theta_W$ close to $1/4$ at
tree-level\cite{Grzadkowski:2006tp,Aranda:2010hn}. In the
phenomenology section, we discuss two possible remedies, by
including either the brane kinetic term
(BKT)\cite{Burdman:2002se,Park:2011an} or an extra $U(1)$ factor,
to address this problem and the consequent modification to the
Higgs mass spectrum.

\section{Group theory analysis}
Following  \cite{Grzadkowski:2006tp}, we adopt the standard
convention for Lie group that  $[H_{i},H_{j}]=0$, $
[\vec{H},E_{\alpha}]=\vec{\alpha}E_{\alpha}$, $
[E_{\alpha},E_{-\alpha}]=\vec{\alpha} \cdot \vec{H}$, and $
[E_{\alpha},E_{\beta}]=N_{\alpha,\beta}E_{\alpha+\beta}$.  Here,
$H$ and $E_{\alpha}$ are the Cartan and root generators of $G_M$
respectively. The structure constant,
 $N_{\alpha,\beta}$,  is given by $
N^{2}_{\alpha,\beta} = n(m+1)(\vec{\alpha} \cdot
\vec{\alpha})/2$~\cite{Lie-algebra}. Moreover, we take the
following normalization for $H$ and $E$:
$\mathrm{tr}H_{i}H_{j}=\delta_{ij}$,
$\mathrm{tr}E_{\alpha}E_{\beta}=\delta_{\alpha+\beta,0}$, and
$\mathrm{tr}E_{\alpha}H_{i}=0~$. The two extra compactified
spatial coordinates, $x^5$ and $x^6$, can be conveniently
described by a complex coordinate $z=(x^{5}+i x^{6})/\sqrt{2}$ and
its conjugate $\bar{z}$. Accordingly, we work with $A_{z}\equiv
(A_{5}-i A_{6})/\sqrt{2}$, the associated gauge field components
in $z$, where $A_{z}=A_{z}^{a}T^{a}$ and $T^{a}$ is the group
generator.

By imposing the proper orbifolding BC's,
 the remaining zero modes of $A_z$, the would-be scalars, are the gauge components associated
with the unbroken group generators $E_{\beta\, ,\,\gamma}$ while
their 4D gauge components possess no zero modes. Unless further
state, the notation $A_z$ is recycled  to collectively signify
these zero modes which carry one  mass dimension,
\begin{equation}
A_{z}=\frac{1}{2}h_{u}E_{\beta}+\frac{1}{2}h_{d}E_{\gamma}
+\frac{1}{2}h_{u}^{\prime}E_{-\beta}+\frac{1}{2}h_{d}^{\prime}E_{-\gamma}~.
\end{equation}
And two would-be-Higgs-doublets can be built up in the following
way:
\begin{equation}
H_1= \frac{1}{\sqrt{2}}\left(  \begin{array}{c} h_u'\\
h_d'\end{array} \right)\,,\;
 H_2= \frac{1}{\sqrt{2}}\left(
\begin{array}{c} -h_d\\ h_u \end{array} \right)\,.
\end{equation}
The SM  $SU(2)_L$ and $U(1)_Y$ groups must be embedded into $G_M$.
If the root vector of the  would-be SM $SU(2)_{L}$  is denoted as
$\vec{\alpha}$, the corresponding  generators read $
J_{0}=\frac{1}{|\vec{\alpha}|^{2}}\,\vec{\alpha}\cdot
\vec{H},~~J_{+}=\frac{\sqrt{2}}{|\vec{\alpha}|}\,E_{\alpha}$, and
$J_{-}=(J_{+})^{\dagger}$.
 The would-be hypercharge generator $Y$ has to be  a linear
combination of Cartan generators and denoted as $\vec{y} \cdot
\vec{H}$. Since the SM group is $SU(2)_L\times U(1)_Y$, hence
$\vec{y}\cdot \vec{\alpha}=0$. The SM  gauge bosons correspond to
the zero modes of the gauge fields associated with these
generators given as,
\begin{equation}
A_\mu= W^+_\mu E_\alpha +W^-_\mu E_{-\alpha} +W^0_\mu
\vec{\alpha}\cdot \vec{H} + B_\mu \vec{y}\cdot \vec{H}.
\end{equation}
In \cite{Aranda:2010hn}, the phenomenologically viable embedding
of the SM electroweak  groups into $G_M$ has been studied, where
the normalization of $\vec{y}$ is fixed such that the SM Higgs
doublet carries hypercharge $1/2$ ($Q=T_3+Y$). The root vectors
$\vec{\alpha}$ and $\vec{y}$ we adopt from \cite{Aranda:2010hn}
are listed in the first two columns in Table \ref{tab:table2}.
From the commutators of $E_{\beta, \gamma}$, one has,
\begin{equation}\label{eq:triangle1}
\vec{\gamma}+\vec{\alpha}=\vec{\beta}\, ~ \mathrm{or}~~ (\,\vec{\beta} - \vec{\alpha} = \vec{\gamma}\,)~.
\end{equation}
Since $h_u$ and $h_d$ transform into each other within one
$SU(2)_L$ doublet, the magnitudes of the two root vectors should
be the same, $|\vec{\beta}|=|\vec{\gamma}|$. Same requirement
applies to the pair of  $h'_u$ and $h'_d$. From
Eq.(\ref{eq:triangle1}), the three root vectors,
$\vec{\alpha},\vec{\beta}$, and $\vec{\gamma}$  form an isosceles
triangle lying on a  plane in the root space. A trivial
geometrical relation follows that
 \begin{equation}
 |\vec{\beta}| \sin \frac{\theta}{2}=\frac{|\vec{\alpha}|}{2}~,
 \label{b-a}
\end{equation}
where $\theta$ is an angle between $\vec{\beta}$ and
$\vec{\gamma}$.
For a simple Lie group, $\theta$ can take only three possible
values: either $\pi/3, \pi/2$, or $2\pi/3$. Hence, the original
group theory problem of embedding the 2HDM in GHU model with gauge
symmetry $G_M$ is now equivalent to looking for the existence of
any equilateral, right isosceles, or obtuse isosceles triangles in
the root diagram of $G_M$.

Once the root $\vec{\alpha}$ is given, one only needs to look up
the corresponding Dynkin diagram and find out which simple root is
adjacent to $\vec{\alpha}$. Next, one looks for the special
isosceles triangle in the 2-dimensional space spanned by the two
simple roots. In Table \ref{tab:table1}, we list all possible
realizations of 2HDM, by employing
Eqs.(\ref{eq:triangle1},\ref{b-a}), in various Lie groups. Note
that all groups, except $G_{2}$ and
 $F_{4}$, have only one possible angle between the two root vectors for the would-be Higgs doublets.

\begin{table}
\caption{ Vertex angles,  isosceles triangles, and candidate
simple Lie groups for the possible realizations of 2HDM.}
\label{tab:table1}
\begin{center} 
\begin{tabular}{|c|c|c|c|}
  \hline
  ~~~$\theta_{\vec{\beta},\vec{\gamma}} $~~~ & ~~~$|\vec{\beta}|$~~~ & ~~~~ candidate groups ~~~ &~~ type of triangle ~~\\
  \hline
  \hline
  $60^{\circ}$ & $|\vec{\alpha}|$                           & $A_{n}, D_{n}, G_{2}, F_{4}, E_{\,6, 7, 8}$ & equilateral  \\ \hline
  $90^{\circ}$ & $\frac{|\vec{\alpha}|}{{\sqrt{2}}}$        & $B_{n}, C_{n}, F_{4}$ &  right  isosceles \\ \hline
  $120^{\circ}$ & $\frac{|\vec{\alpha}|}{{\sqrt{3}}}$       & $G_{2}$             & ~obtuse isosceles \\ \hline
\end{tabular}
\end{center}
\end{table}
\begin{figure}
\begin{center}
\includegraphics[height=0.17\textheight,width=0.22\textwidth]{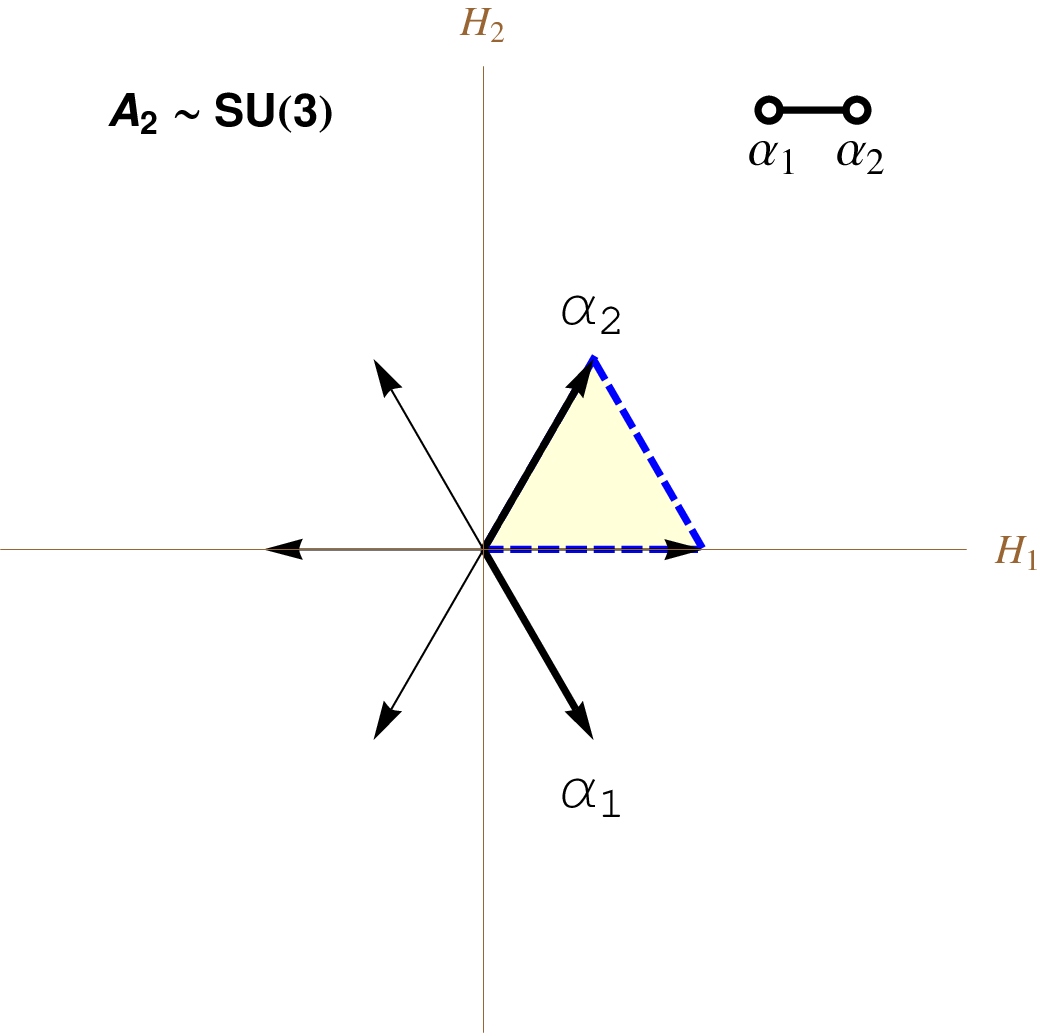}
\includegraphics[height=0.17\textheight,width=0.22\textwidth]{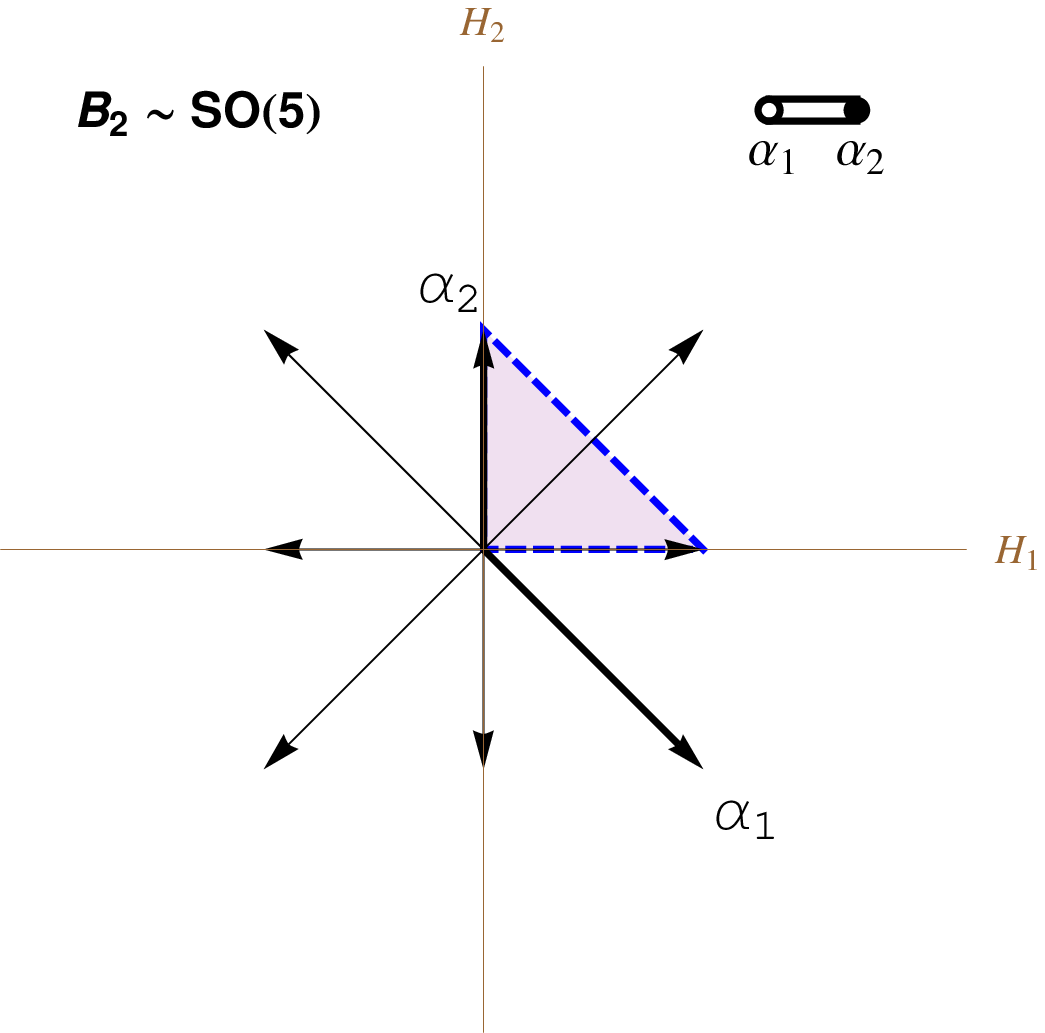}
\includegraphics[height=0.17\textheight,width=0.22\textwidth]{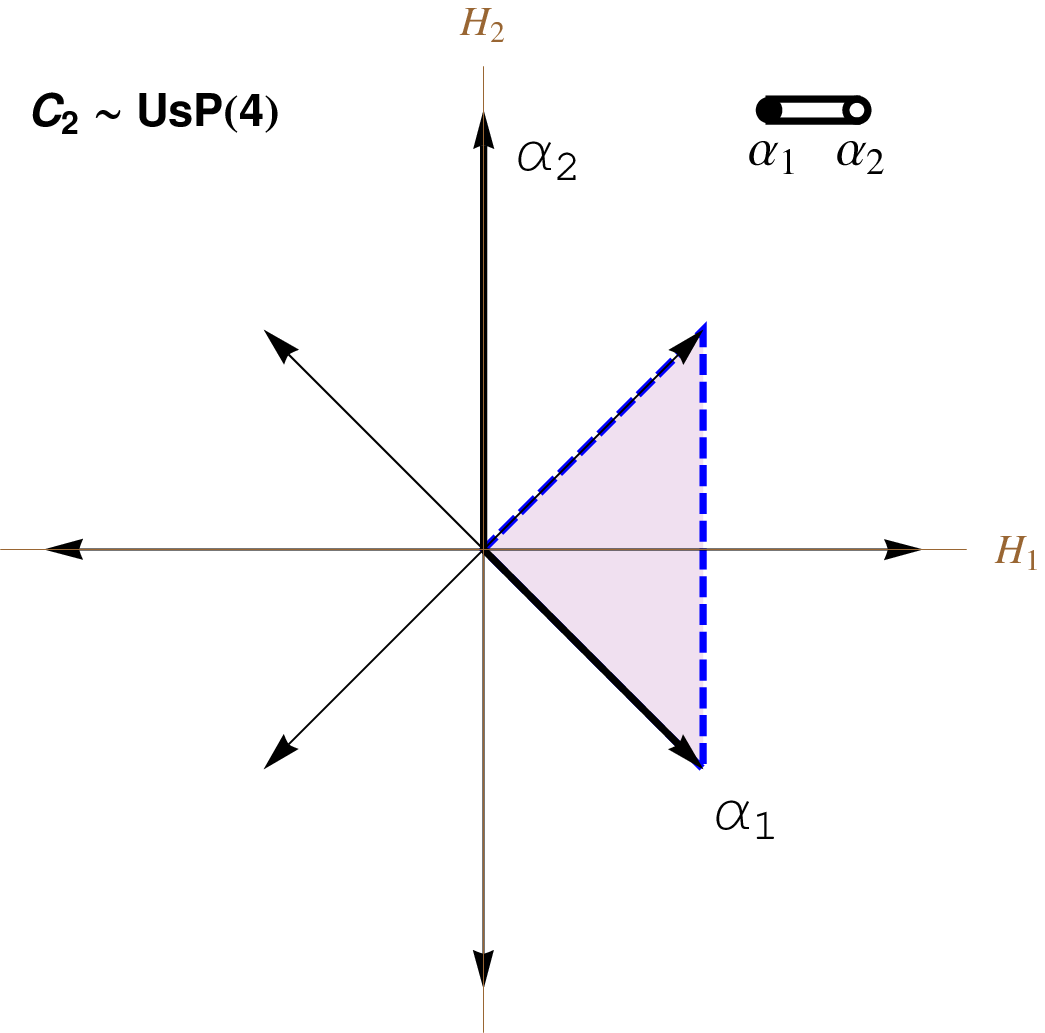}
\includegraphics[height=0.17\textheight,width=0.22\textwidth]{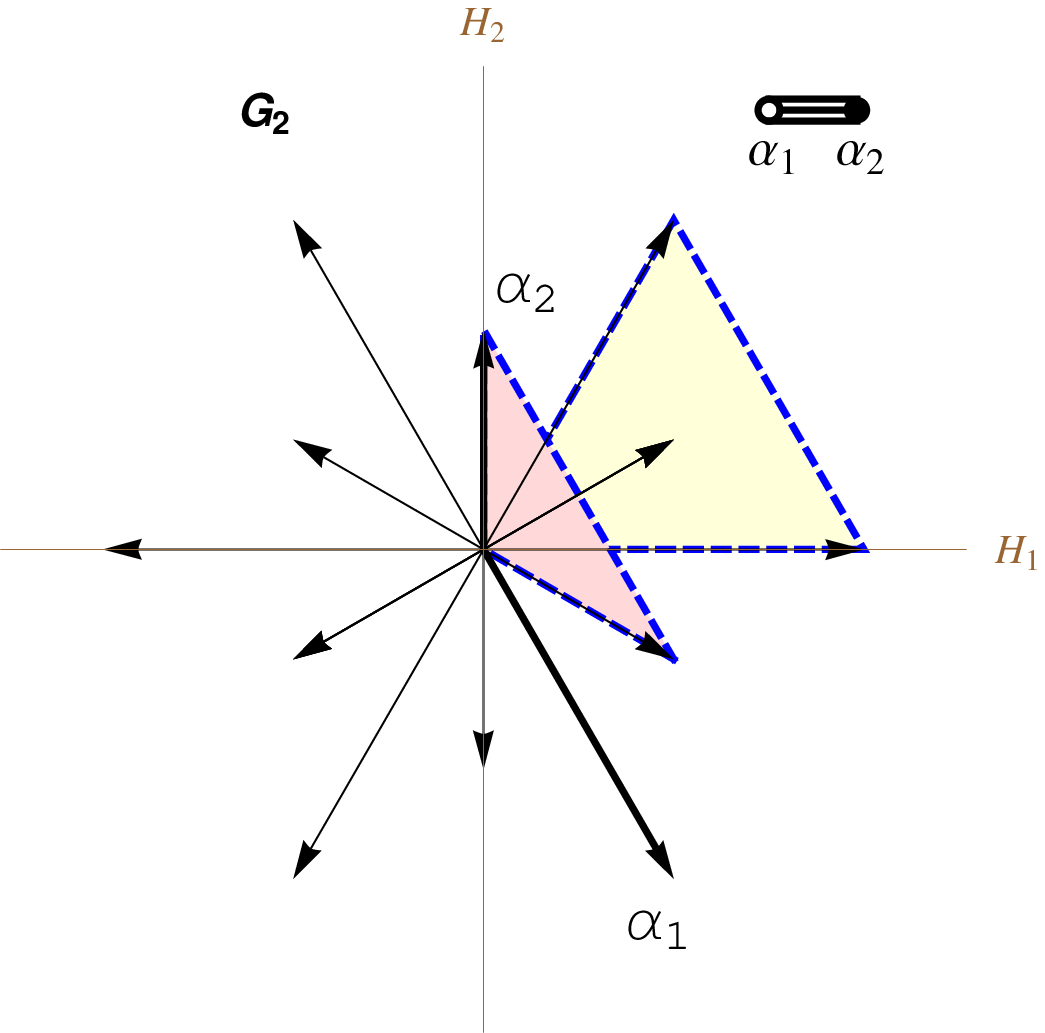}
\caption{Root diagrams, Dynkin diagrams and corresponding triangles of $A_{2}$  (left upper), $B_{2}$ (right upper), $C_{2}$ (left bottom) and $G_{2}$ (right bottom) where yellow, purple and red triangles represent the equilateral, right isosceles and obtuse isosceles triangles, respectively.
}\label{fig:figure1}
\end{center}
\end{figure}
We illustrate the finding by four rank-2 groups,
 $A_2, B_2, C_2$, and $G_2$, which can be
diagrammatically summarized in the self-explanatory
Fig.\ref{fig:figure1}.  Here, the triangles formed by the
corresponding root vectors, $\vec{\alpha},\vec{\beta}$, and
$\vec{\gamma}$, are highlighted by color shades. Note that there
are two distinct isosceles triangles can be drawn with
$\vec{\alpha}$ in the root diagram for $G_2$. Therefore, there are
four possible forms for $V_4$ (two reds, two yellows, red with
yellow on one side, and  red with yellow on the opposite sides) in
the $G_2$-based GHU model.
 Since the root $\vec{\alpha}$ for SM $SU(2)_L$ is always
at the end of the Dynkin diagram, the analysis for $G_M$ with rank
higher than two is no more complicated than those displayed in
Fig.\ref{fig:figure1}.

\section{Higgs potential and weak mixing}
The  scalar quartic coupling terms in 2HDM
arise from the 6-dimensional gauge field strength square,
$g_{4D}^{2}\,\mathrm{tr}[A_{Z},A_{Z}^{\dagger}]^2$,
\begin{eqnarray}
 V_4 &=& \frac{g_{4D}^{\,2}\vec{\beta}^{\,2}}{4}\Bigg\{
                                   \Big( |h_{d}|^{2} -|h_{u}^{\prime}|^{2}\,\Big)^{2}  +\Big( |h_{u}|^{2} + |h_{d}^{\prime}|^{2}\,\Big)^{2}\nonumber \\
                    && +2(\tilde{N}_{-\gamma,\beta}^{2}+1)\Big(|h_{u}|^{2}|h_{d}|^{2} + |h_{u}^{\prime}|^{2}|h_{d}^{\prime}|^{2}\Big)\nonumber \\
                               && +2(\tilde{N}_{\gamma,\beta}^{2}-\cos\theta)\Big (|h_{d}|^{2}|h_{d}^{\prime}|^{2} + |h_{u}|^{2}|h_{u}^{\prime}|^{2}) \nonumber \\
                              &&+2\Big(\tilde{N}_{-\,\gamma,\beta}^{2} + \tilde{N}_{\gamma,\beta}^{2} \Big)
                                (h_{d}h_{u}^{*}h_{u}^{\prime}h_{d}^{\prime *} + h_{d}^{*}h_{u}h_{u}^{\prime *}h_{d}^{\prime})
                                            \Bigg\}~,\label{quartic}
\end{eqnarray}
\normalsize where $\tilde{N}_{\gamma,\beta}^{2}\equiv
N_{\gamma,\beta}^{2}/{\vec{\beta}^{\,2}}$. Also,
Eqs.(\ref{eq:triangle1},\ref{b-a}) and the relations
$N_{\gamma,\beta}=-N_{\beta,\gamma}=-N_{-\alpha,-\gamma}~$ have
been used to arrive in the above expression. Unsurprisingly, the
resulting Higgs potential is completely determined by the gauge
group $G_M$ for a given root vector $\vec{\alpha}$ for SM
$SU(2)_L$.

Eq.(\ref{quartic}) can be written in a much more compact form as:
\begin{eqnarray}
V_4(H_{1},H_{2})=
\frac{1}{2}\lambda\,\Big(|H_1|^{2}-|H_{2}|^{2}\Big)^2
+ \frac{1}{2}N\lambda\,|H_{1}^{\,\dagger}H_{2}|^2~, \label{eq:GHUpotential}
\end{eqnarray}
where $\lambda=|\vec{\beta}|^2 g^2_{4D}$ and $N$ is an integer
depending on the group $G_M$,  as shown in  Table
\ref{tab:table2}. In  Eq.(\ref{eq:GHUpotential}),  $H_1$ and $H_2$
have the desired hypercharge either $Y=1/2$ or $Y=-1/2$.
Geometrically speaking, two identical or the mirror pair  root
triangles  are adopted for the two would-be Higgs doublets. The
cross coupling  between $H_1$ and $H_2$, $\frac{1}{2}N\lambda$,
can be calculated from the ladder chain of root vectors. For $G_2$
and $F_4$, there are two possible root triangles that can be
adopted for 2HDM. For each realization, $\vec{y}$ has to be
normalized accordingly, thus two possible $N$'s for $G_2$ and
$F_4$. Although the origins are totally different,
Eq.(\ref{eq:GHUpotential}) is accidentally identical to the
$D-$term Higgs potential in the Minimal Supersymmetric SM(MSSM) if
one substitutes $\frac{1}{2}\lambda \Longrightarrow
\frac{g^{2}+g^{\prime\,2}}{8}$, and $ \frac{1}{2}N\lambda
\Longrightarrow \frac{g^{2}}{4}$. It is thus expected that the
2HDM in GHU and the MSSM share a similar physical Higgs mass
spectrum.

Among all the Lie groups, the $G_{2}$-based GHU models have the
richest 2HDM phenomenology. In addition to the quartic coupling
terms given in  Eq.(\ref{eq:GHUpotential}), the 2HDM based on
$G_2$ can have two extra possible forms for $V_4$:
\begin{eqnarray}
&& V_4(H_{1},H_{2}) =
\frac{1}{2}\lambda\Big[\,\frac{1}{3}(H_{1}^{\dagger}H_{1})^{2}+(H_{2}^{\dagger}H_{2})^{2}
\nonumber\\
&& +N_{1}(H_{1}^{\dagger}H_{1})(H_{2}^{\dagger}H_{2}) +
N_{2}(H_{1}^{\dagger}\sigma^{a}H_{2})(H_{2}^{\dagger}\sigma^{a}H_{1})
\Big]~, \label{eq:G2potential}
\end{eqnarray}
 where ($N_{1}$, $N_{2}$) is either $(-2,+1)$ ( as
discussed in ~\cite{g_2}) or $(+4,-2)$, corresponding to the
yellow and red triangles in Fig. \ref{fig:figure1} on the opposite
sides or on the same side respectively. For $G_2$, $\vec{y}$
aligns with the medians of two possible root triangles. Hence only
one of the two Higgs doublets can acquire the desired hypercharge.
In Eq.(\ref{eq:G2potential}), $H_1 (H_2)$ carries hypercharge $1/2
(3/2)$.  Therefore, only $H_1$ is responsible for the  EWSB, and
$H_2$ is not the canonical Higgs doublet with $|Y|=1/2$. We note
in passing that since the two possible root triangles of $F_4$ lie
on different planes,   not so interesting $V_4$ with
$(N_1,N_2)=(0,0)$ can emerge.

\begin{widetext}
\begin{center} 
\begin{table}[h!t]
\caption{ The candidate simple Lie groups, based on which the 6
dimensional GHU model is phenomenologically
viable~\cite{Aranda:2010hn},  their root vectors $\vec{\alpha}$
and $\vec{y}$ for $SU(2)_L$ and $U(1)_Y$ respectively, and all
other relevant numbers, see text.  Here, $\alpha^k$ is the $k-$th
simple root labelled by the Dynkin diagram, and $\tilde{\mu}_k$ is
the rescaled $k-$th fundamental weights such that
$\vec{\alpha}^i\cdot \tilde{\mu}_j=\delta_{ij}$. Note that the SM
quark representation cannot be accommodated in the GHU models
based on $C_n$ or $D_n$ group ~\cite{Aranda:2010hn}. However $C_n$
and $D_n$ are listed here for the sake of comparison and
completeness.
 } \label{tab:table2}
\begin{center} 
\begin{tabular}{|c||c|c|c||c|c|c||c|}
  \hline
  \hline
 Group & $\alpha$ & y & $\tan \theta_{W}$ & $~~~\tilde{N}^{2}_{-\gamma,\beta}~~~$ & $~~~\tilde{N}^{2}_{\gamma,\beta}~~~$ & $~~~\theta_{\beta,\gamma}~~~$ & $~~~N~~~$ \\
  \hline
  \hline
  $A:SU(3l)$ & $\alpha^{1}$ & $\tilde{\mu}_{2}/2$ & $\sqrt{3l/(3l-2)}$ & $1/2$ & 0 & $60^{\circ}$ & 1 \\
  \hline
  $B:SO(2n+1)$ & $\alpha^{1}$ & $\tilde{\mu}_{2}/6$ & $\sqrt{3}$ & $1$ & $1$ & $90^{\circ}$ & 4   \\
  \hline
  $C:USp\,(2n)$ & $\alpha^{n}$ & $\tilde{\mu}_{n-1}/2$ & $\sqrt{1/(n-1)}$ & $1$ & $1$ & $90^{\circ}$ & 4\\
  \hline
  $D:SO(2n)$ & $~\alpha^{1}, \alpha^{n,~ n-1}~$ & $~\tilde{\mu}_{2}/2$, $\tilde{\mu}_{n-2}/2~$ & $\sqrt{2/(n-1)}$ & $1/2$ & 0 & $60^{\circ}$ & 1 \\
  \hline
  $G_{2}$ & $\alpha^{1}$ &  $\tilde{\mu}_{2}/2$, $\tilde{\mu}_{2}/6$ & $\sqrt{1/3}$, $\sqrt{3}$
  & $1/2 $, $3/2$ & $0$, $2$ & $ 60^{\circ}$, $120^{\circ}$&   1, 7 \\
  \hline
  $F_{4}$ & $\alpha^{1}$ & $ \tilde{\mu}_{2}/2$, $\tilde{\mu}_{2}/6$   & $ \sqrt{1/3}$, $\sqrt{3}$  & $1/2 $, $1$ & $0$, $1$ & $ 60^{\circ}$, $90^{\circ}$  & 1, 4\\
  \hline
  $E_{6}$ & $\alpha^{1,5}$ & $\tilde{\mu}_{2,3}/2$ & $\sqrt{3/5}$ & $1/2$ & 0 & $60^{\circ}$ & 1\\
  \hline
  $E_{7}$ & $\alpha^{1,7}$ & $\tilde{\mu}_{2,3}/6$ & $\sqrt{3}$,$\sqrt{3/2}$ & $1/2$ & 0 &  $60^{\circ}$ & 1\\
  \hline
  $E_{8}$ & $\alpha^{1,8}$ & $\tilde{\mu}_{2,3}/6$ &$\sqrt{9/7}$,$\sqrt{3/5}$ & $1/2$ & 0 &  $60^{\circ}$ & 1\\
  \hline
\end{tabular}
\end{center}
\end{table}
\end{center}
\end{widetext}

As  mentioned in the introduction, the Higgs quadratic couplings
in GHU model must be generated radiatively by some
symmetry-breaking mechanism~\cite{Antoniadis:2001cv}. The
resulting quadratic couplings are highly model dependent.
Here, we take a bottom-up approach and treat all quadratic couplings as
phenomenological parameters. Then the full $SU(2)$ invariant
scalar potential of 2HDM reads,
\begin{equation}
V= m_1^2|H_1|^2+m_2^2
|H_2|^2-m^2_{12}(H^{\dagger}_1H_2+H^{\dagger}_2H_1 ) +V_4 ~.
\end{equation}
Based on this, a textbook EWSB analysis can be performed
straightforwardly. We denote the vacuum expected values as
$\langle H^0_1 \rangle =v_1$ and $\langle H^0_1 \rangle =v_2$ with
$\sqrt{v_1^2+v_2^2}\sim 246$ GeV, and $\tan\beta = v_2/v_1$. The
masses of CP odd pseudo-scalar $A_0$ and charged Higgs $H^{+}$ are
given by $M^2_{A_0}=m^2_{12}/\sin\beta\cos\beta$ and
$M^2_{H^{+}}=M^2_{A_0}-\frac{1}{4}N\lambda v^2$, respectively.
Similar to MSSM, the tree-level mass of the lightest neutral
Higgs has an upper bound, $M_h\leq \sqrt{\lambda
v^2}(\sqrt{7\lambda v^2}/2)$ for N=1 and 4 (N=7)
, which may
causes a phenomenological problem.
However, just like in MSSM,
there are many heavy degrees of freedom beyond SM in the GHU
model.
Similar to the positive $\delta M_h$ due to the stop loops in the MSSM \cite{Okada:1990vk},
it is well known that the
inclusion of all the radiative corrections from either the bulk fields and their KK excitations, the brane-localized
fields, or the brane kinematic terms could largely enhance $M_h$ from the above tree-level prediction\cite {Antoniadis:2001cv}.
However, a general discussion on this issue is still lacking.
Such a model-independent discussion on the radiative corrections to $M_h$ is beyond the scope of the present paper,
here we just assume that a careful consideration which includes
the radiative corrections to $V_4$ can rescue the lightest neutral
Higgs mass problem as in MSSM. The investigation along this line
will be presented elsewhere \cite{next}.

Finally, we address the weak mixing angle problem in GHU which is
independent of $d$. Adding the BKT to GHU model is one of the
remedies to obtain a realistic $\sin^2\theta_W$ close to
experimental value. We use the following BKT which involves gauge
zero modes on the brane only, \small
\begin{equation}
\mathcal{L}_{B.K}=  -\frac{1}{4}\, \int d^{4}x  dx^5 dx^6
~\delta(x^5)\delta(x^6) \Big[ c_{1}(F_{\mu\nu}^{(0)\, a})^{2} +
c_{2}(F_{\mu\nu}^{(0)\, b})^{2} \Big]
\end{equation}
\normalsize where the superscript $(0)$ represents zero mode,
$c_{1}$, $c_{2}$ are free parameters with mass dimension -2, and
$a,b$ are the group indices for SM $SU(2)_L$ and $U(1)_Y$,
respectively. The 4D effective gauge couplings  are modified:
\begin{equation}
g^i_{4D} \rightarrow \frac{g^i_{4D}}{\sqrt{Z_{i}}},
~~~\mbox{with}~\mathrm{Z}_{i}=1+\frac{c_{i}}{\mathrm{Z}_{0}^{2}}~,~~\mathrm{(i=1~or~2)}~,
\end{equation}
where $Z^2_0$ is the volume of the two extra dimensions. As a
result,  the weak mixing angle becomes
\begin{equation}
\tan \theta_{W} \rightarrow \sqrt{\frac{Z_{1}}{Z_{2}}}\,{\tan
\theta_{W}}~.
\end{equation}
From the  experimental value  $\tan \theta_{W}
=0.5356$~\cite{Nakamura:2010zzi},  the ratio of $Z_1/Z_2$ can be
fixed for a given $G_M$. Moreover, the upper bound of the lightest
neutral Higgs mass gets a factor of  $\sqrt{Z_1}$ enhencement and
this problem can be alleviated as well.

Adding an additional $U(1)^{\prime}$ gauge group is an alternative
to obtain a realistic weak mixing angle in GHU model. The mixing
between SM $U(1)_Y$ and $U(1)^{\prime}$ leads to an effective
$Z_2\neq 1$ ( with $Z_1=1$)  as in the previous case. The extra
$U(1)$ factor introduces a new electrically neutral gauge boson
whose mass shall be arranged to be the order of the
compactification scale. The original $\tan\theta_W$ for each group
without $U(1)^\prime$ can be found in table \ref{tab:table2}. Note
that to obtain the realistic $\theta_W$, $SU(3), SO(2n+1), G_{2},
F_{4}$ and $E_{7}$ require the largest $Z_2$ among all the groups.

Before concluding, we remark on the new scalar boson recently observed near $126$ GeV with the diphoton excess
at the  large hadron collider (LHC)~\cite{:2012gk, :2012gu}.
Although most of its physical properties  seem to be
consistent with the elementary Higgs boson in the SM, the diphoton excess may indicate the existence of
new physics beyond the SM. As discussed, $M_h\sim 126$ GeV  could be easily accommodated in a realistic 2HDM in the GHU models
with some construction dependent extensions.
We note in passing that the diphoton excess, if persist, could also be explained in this GHU framework
for there are many charged  heavy degrees of freedom.

In summary, we perform a general group theory analysis on the
realization of 2HDM in GHU model with a 6-dimensional gauge $G_M$
symmetry, where $G_M$ is a simple Lie group. We showed that a 2HDM
at low energy can possibly be made if  the three root vectors
associated with the would-be Higgs doublets and the SM $SU(2)_L$
form isosceles triangles with vertex angle either of $\pi/3,
\pi/2$ or $2\pi/3$. The quartic coupling terms in the 4D effective
2HDM potential are completely determined by the group $G_M$ at
tree-level. Only a few potential forms can be admitted for all
possible Lie groups,   as shown in  Table \ref{tab:table2}.
Moreover, which form to be admitted depends solely on the vertex
angle. We observed that, among all possible Lie groups, the 2HDM
based on the GHU model with $G_2$ gauge symmetry has the richest
structure in the Higgs potential. We discuss the mass
spectrum of physical Higgs bosons and two possible remedies to
obtain a realistic weak mixing angle in GHU models as well.
Finally, we briefly comment how to accommodate the recent observed $\sim 126$ GeV scalar boson
at the LHC in this GHU framework.
\\ \\ \\
\begin{acknowledgments}
W.F.C. was supported by the Taiwan NSC, Grant No.
99-2112-M-007-006-MY3. S.K.K  was  supported by NRF and MEST of
Korea,  No. 2011-0029758. J. P. was supported by the Taiwan NSC,
Grant No. 100-2811-M-007-030 and 099-2811-M-007-077. J.P. thanks
D.W. Jeong for  valuable comments and discussion.
\end{acknowledgments}

\end{document}